# TOLLAN-XICOCOTITLAN: A RECONSTRUCTED CITY BY AUGMENTED REALITY (EXTENDED)

M. en C. Martha Rosa Cordero López, M. en C. Marco Antonio Dorantes González.

Escuela Superior de Cómputo, I.P.N,
México D.F.
Tel. 57-29-6000 ext. 52065 y 52032.

E-mail: mcorderol@ipn.mx
mdorantesg@ipn.mx

## *ABSTRACT*

*This project presents the analysis, design, implementation and results of Reconstruction Xicocotitlan Tollan-through augmented reality, which will release information about the Toltec culture supplemented by presenting an overview of the main premises of the Xicocotitlan Tollan city supported dimensional models based on the augmented reality technique showing the user a virtual representation of buildings in Tollan.*

## *KEYWORDS*

*Augmented Reality, Databases, Visual Programming, Virtual Reconstruction, Archaeological Site*.

## 1. INTRODUCTION

The Archeological Zone of Tula, is the most important of Tolteca Culture. It's conformed by a set of buildings with a religious symbolism, for example the Central Altar, the Coatepantli (wall of Snakes), Burnt Palace, ball games and the Tzompantli. Thanks for the National Institute of Anthropology and History (INAH) opened in Tula a museum about Tolteca Culture.

Thanks to science and technology have made great discoveries and changes in society over time [1]. Nowadays, it is possible to combine virtual and real objects within the same environment, to create supplemented views from somewhere that people are viewing [2]. This process is called Augmented Reality (AR) [3] [4].

The project applies AR together with archaeological knowledge of the Tollan-Xicocotitlan city, in Tula, Hidalgo. In order to obtain a system that models projecting three-dimensional (3D) showing the architecture of the buildings constructed there and complemented with written information about each campus.

Indeed, the reconstruction allows displaying any building that is in ruins, presenting it in three-dimensional model of the structure information. Besides the system provides support in order to have better idea of the constructed buildings in the past. The system can be applied in various places, with desired display information from the Toltec culture, a museum, exhibition or educational institutions where they are taught subjects related to the teaching of the Hispanic cultures.

AR supports markers located on a fixed surface, such as the ruins of a temple, a pyramid or a display in a museum. Such markers are detected by the input devices that should be placed in a specific position for the mark to be recognized and to be viewed on virtual model for the whole environment. The viewer appreciates a virtual city by means of the system which builds boom in architectural, or reconstruction of events occurred in the past.

As a result, AR does not absolve the user from the reality, all experiences become more interesting for visitors, who are immerse in a particular event occurred in the past.

## 2. PRINCIPLE

**Analysis**

To construct a model system the engineer would consider some restrictions:

a) **Postulation:** It reduces the number of permutations and possible variations to reflect the problem in a rational way by the model.
b) **Simplifications:** It creates the model on time.
c) **Limitations and restrictions:** It helps to delimitate the system and guides the way to create the model also the approach it takes to implement the model.

Based on the above definitions the restrictions on our system are described below:

**Postulation:**

- The system should store three-dimensional models of the main buildings in the city of Tollan-Xicocotitlan.
- The system displays information submitted buildings.
- The system offer the opportunity to comment on the experience in using application, also may make recommendations for improving the application.
- There are a manager who are watching the publications made in the system.

**Simplifications:**

- Only the main venues of the city are modeled.
- Only relevant and concise information on the site submitted by the visitor are present.

**Scopes:**
- Visitors can view the graphic images from their different perspectives.
- The system may be entered in any display outside the archaeological site, allowing more people to know about the Toltec culture.
- The system could be adapted to real buildings that make Tollan-Xicocotitlan appreciated.

**Actors involved in the requirements definition.**

- **Archaeologists:** With the knowledge we have about the archaeological site, the Archaeologists help to define the project providing information on main buildings of the Tollan-Xicocotitlan City.

- **Visitors.** Visitors can share concerns and attractive to consider important on the site and can add visual and operational aspects they would like to see in the system.

**Technical environment of the system or product developed.**

An application or modeler is used to generate models of the main buildings, libraries that allow recognition of patterns Augmented Reality and a programming IDE for the conjunction of the elements used.

**Requirements analysis.**

The requirements analysis is one of the most important tasks in the life cycle of software development; it sets the planning of the application.

The requirements analysis can be defined, as the process of studying the user needs to get a definition of the system requirements, hardware or software, and the process of study and refinement of these requirements, definition provided by the IEEE [10]. Also a required is defined as a condition or ability that the user needs to solve a problem or achieve a determined goal [10]. This definition extends and applies to the conditions to be met or have a system or a component to satisfy a contract, standard or specification.

Based on the above definitions have been found the following list of functional requirements for the present system:

- **Main Functional Requirements**

| ID | Functional Requirements | Description | Priority |
|---|---|---|---|
| RF1 | Visualize the building focused. | Allow the user to view a reconstruction | High |
| RF2 | Show relevant dates from the building. | It shows to the visitor the information building. | Medium |
| RF3 | Recognize the A.R. mark that is set to each building. | To load a virtual model, it is important to recognize the mark assigned to each building. | High |
| RF4 | Guarantee the right over position of the virtual objects. | The system should provide the ideal model for each virtual | High |

Based on the functional requirements we begin a technique using the UML system, which is below.

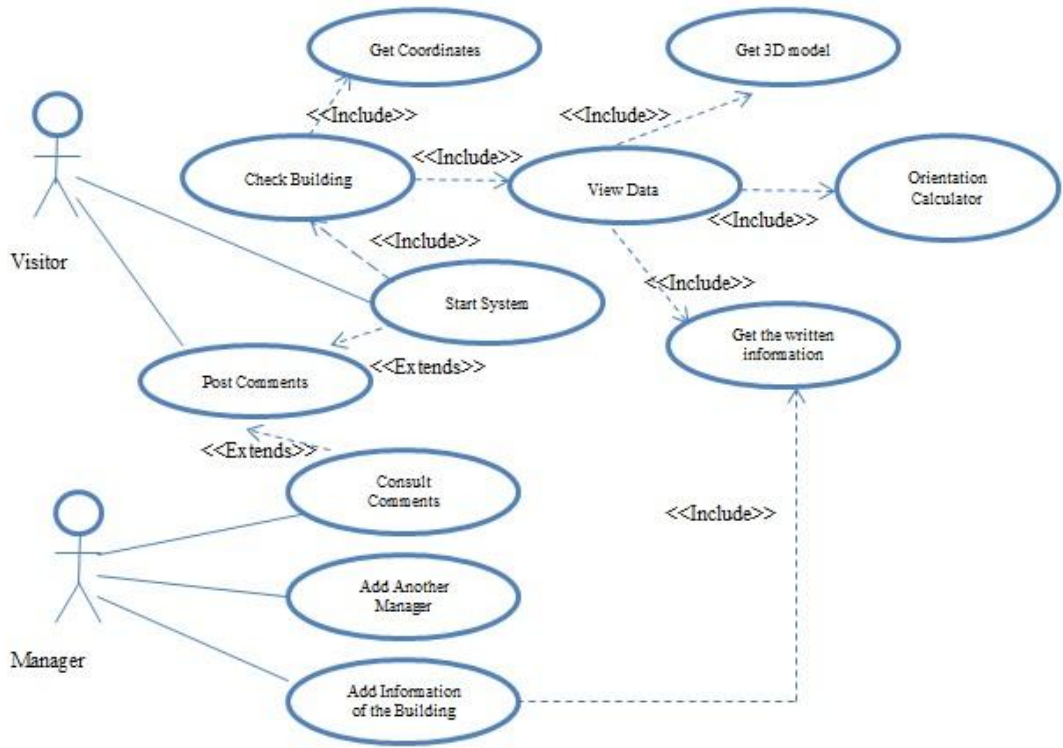

**UML diagram description**

Use Case Start System
- Based on functional requirements the visitor is the one who starts the system's tour. The system needs to be abled and the local network has to be available. The system displays the welcome screen to visitors, this screen contains the following:
    o Start the system.
    o Post comments.
- The visitor selects an option and depending on that, the system will display the relevant information.

Use Case Post Comments
- The visitor could post comments after selecting the generate report option.

Use Case Check Buildings
- The system starts capturing video images based on the objects which are focused with the camera and the system displays the images received from it. If the system detects an Augmented Reality mark associated to an object that has been approached by the actor the use case includes the functionality of the Use Case "Getting coordinates" and waits for a response back, receiving data from the coordinates of the 3D object.

- The system sends the data received from the previous step to the use case "View data" to present three-dimensional models to the user.
- The system receives the data from the previous step and presents the user with three-dimensional model associated with the detected Augmented Reality mark.

Use Case View Data

- The system receives the coordinates and sends a parameter with the name to get the associated 3D model with the coordinates.
- The received information is processed to be presented on screen.
- The name of the building is sent to present the relevant information of the building and the information is received and it fits to be presented on the screen.
- To determine the angle and the screen position of the building it uses the Use Case 'Orientation calculator', then the generated information is received and the virtual model with the associated information of the building is presented on screen.

Use Case Orientation Calculator

- The system implements operations of rotation, translation and scaling matrices on the three-dimensional model in order to match the angle members.
- After the angle has been equaled the system returns an array of the position values of the virtual object with respect to the viewing angle of the visitor.

Use Case Get Coordinates

- The system detects an Augmented Reality mark associated with an object that has been focused with the camera and the system performs a comparison between the detected marks and the stored marks in the database to determine which 3D model will be called.
- The system sends the data from the comparison to the use case 'Check Building'.
- 

Use Case Add Information

- The manager adds information of the Building to the system.

Use Case Add Another Manager

- The manager adds another manager if it is necessary.

Use Case Consult Comments

- The manager can consult comments that have been posted in the system.

**Construction of the application**

The visitor should focus the camera to the A.R. marks, and a frame captures a real world image through a camera.

The image is modified taking into an account a certain threshold value. Thus, the pixels whose intensity exceeds the threshold are converted into white pixels. The remainder is transformed into black pixels then black frames are sought and found as the existing marks in the image.

If the shape of the analyzed mark and the stored mark match, it is used the size information and orientation of the stored mark to compare with the mark that has been detected in order to calculate the position and orientation of the camera relative to the mark, and stored it in an array

Then the matrix establishes the position and orientation of the virtual camera (processing the camera view), equivalent to a transformation of the object coordinates to draw. Once we put the virtual camera in the same position and orientation as the real camera, the virtual object is drawn on the mark, the resulting image is displayed, containing the image of the real world and the virtual object superimposed, aligned on mark. It performs the same process with the following frames.

**Augmented reality**

AR techniques add virtual elements to real world, scientists have been constructing prototypes, the first was created for Ivan Sutherland, it was in 1960 when he used a 3D images display device to see 3D graphics, lately in 1962 the "Sensorama" was created for Morton Heilig.

Augmented Reality can be used in:

- **Scholar projects:** They are used in museums, exhibitions or thematic parks due to the price isn't enough to be improved in the domestic area.
- **Simulation:** It's used to simulate flights, and land paths, or military entrainment.
- **Emergency services:** In case of emergency, Augmented Reality can show instructions of what to do to evacuate the place.

**Operation of augmented reality**

Three basic key elements of AR are:

- Display (output),
- Location of virtual objects in the real world (registration),
- Methods interaction (input).

The main point in the development of an AR application is a motion tracking system. RA technique relies on "Bookmarks" or an array of markers within the field of view of the cameras, such that a computer system has a benchmark on which superimpose images. These markers are predefined by the system and the pictogram can be unique for each image to be superimposed or simple shapes, such as picture frames, or textures within the field of view.

A computer system can be more intelligent, able to recognize simple shapes, such as the floor, objects like chair, table, and simple geometric shapes, to name a cell phone on the table that can be used with a mark or even with the human body that can be used with the same purpose.

The following figure shows an example of the marks described in the previous paragraph.

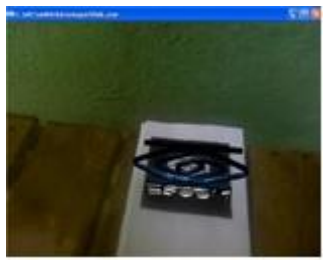

**Fig. 1 Final Result of A.R. using a mark and a model.**

# TOOLS

**ARToolKit**

ARToolKit is a set of libraries for C / C + +, that are useful for building AR applications. It includes a number of computer vision techniques for video capture and pattern searching for capturing images.

Users believe that only in the real world it is possible to perform transformation on objects. But, we want to show that it is possible to perform this kind of transformation on virtual objects. Users are able to see this transformation via the camera or by capturing them, taking into account position, size, orientation, and lighting, as these objects would be perceived by the user in the real world, if they were actually there.

Once a template is detected within an image, studying the orientation, position, and size of the template, the application is able to calculate the relative position and orientation of the camera, and relative to the template. Using this information, you can draw the corresponding object on the captured image by means of the ARToolKit external libraries (e.g., GLUT and OpenGL).

In this way, the object appears on the template, in the position, orientation, and size corresponding to the view taking by the camera. Due to the number of possibilities are big, the application take a decision to select one, taking into account the information of other various operations.

**Nyartoolkit**

NyARToolkit provides a trail marker based on AR. However, the software has been optimized for easy portability among different programming languages. In order to develop an application running AR on different platforms and operating systems, NyARToolkit libraries are the best option.

NyARToolkit include some key features, like:

- Bookmarks AR based tracking.
- Support for desktop and mobile platforms.
- Scoreboard optimized and enhanced survey.

**Blender**

Blender is a tool for creating mainly modeling animation and creation of three-dimensional graphics. Some features are:

- It is a cross-platform tool, is free software and complies with the functionality provided similar commercial tools.
- Along with the animation tools including inverse kinematics, armature or grid deformations, loading and particle vertices static and dynamic.
- Features interactive games such as collision detection, dynamics and logic recreations

## 3. EXPERIMENTAL RESULTS

**Development**

The system is divided into two main modules User(Visitor) and Manager.

The user(Visitor) module is in charge for presenting 3D models of each building, in this module, users can visualize a pyramid in 3D and can also comment on the experience that let them use this type of system.

The manager module allows manager to upload new handling system, just as you can modify the information associated with each building, this section manager perform the query of comments made by users of the system.

**Main menu**

In the main menu you can access it as administrator, see the buildings and post a comment.

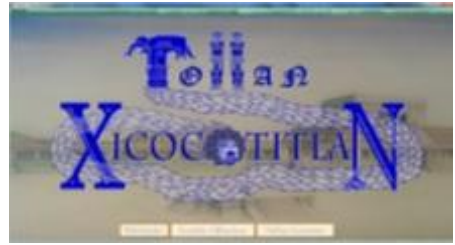

**Fig. 2 Main menu of the system**

**Post a comment**

If the users add the requested data by the system they could post a comment.

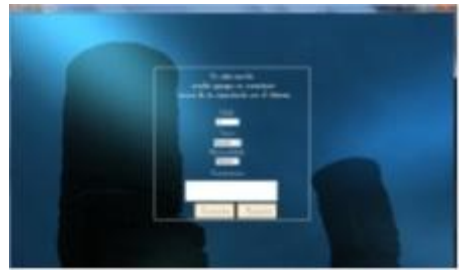

**Fig. 3 Post a comment in the system**

**Manage Menu**

The manager could select several options such as consult comments, add new managers, change his own information and the buildings presented in the system.

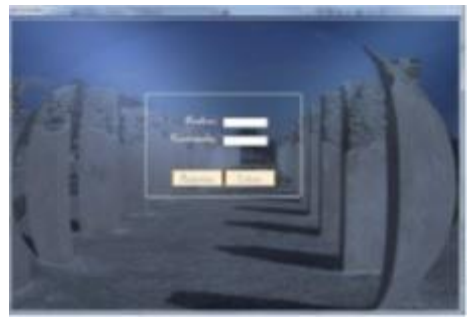

**Fig. 4 the manage menu**

**Creation of marks**

We have generated our own marks to system purposes, which will allow to over put the 3D models that will be showed to the final user. The map below shows the marks and 3D buildings of Tollan City.

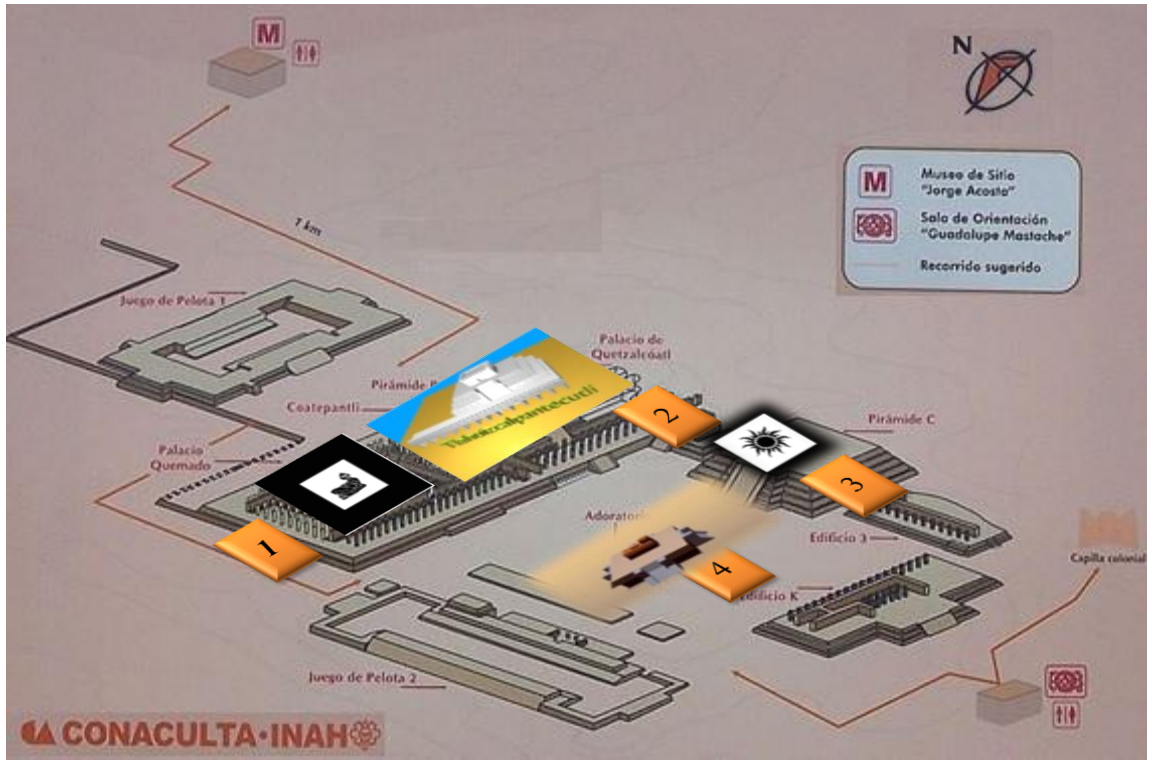

**Fig. 5 The Tollan-Xicocotitlán map [11].**

1. Represents the proposed mark to burnt palace.
2. Represents a proposed 3D view of Pyramid B.
3. Represents the proposed mark to pyramid C.
4. It shows the 3D view of the Shrine

**Examples of Blender Modeling**

We have made some 3D models of the Tollan City. The following images have been developed in Blender:

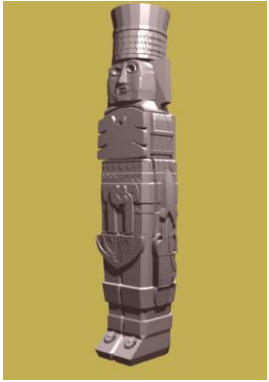

The "Atlantes" are assembled sculptures with four parts and were a "Toltecas" expression of art. These sculptures represent warriors with feather crowns, carrying weapons such as knives, arrows and throws darts (atlas). [12]

In the Fig. 21 we can see a 3D model of an "Atlante" developing in Blender.

**Fig. 6 3D Model of "Atlante de Tula"**

We can see the reconstruction of the Burnt Palace in a 3D model developed with Blender at the Fig 22. Also we can see the reconstruction of Tollan City at the Fig. 23.

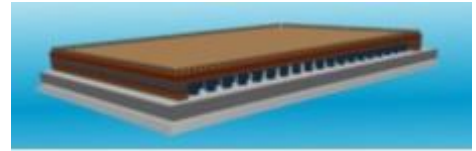

**Fig. 7 Burnt Palace.**

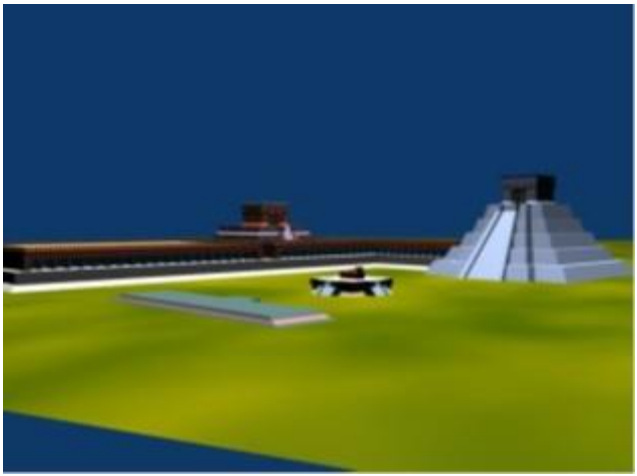

**Fig. 8 Tollan Xicocotitlan View.**

**Tests with ARToolKit**

After understanding the operation of the libraries were established own marks of AR and generated three-dimensional models that would be superimposed on these markers. The results are shown below:

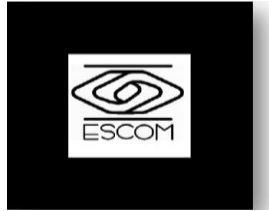 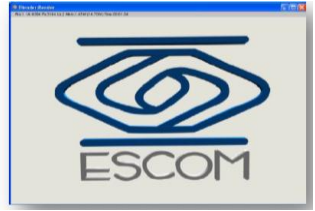

**Fig. 9 Mark of A.R.**          **Fig. 10 Three Dimensional model**

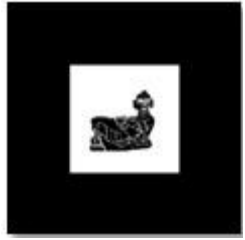 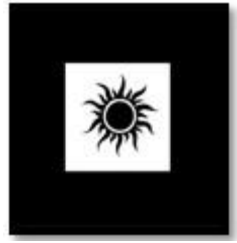

**Fig. 11 Mark of Burnt Palace**   **Fig. 12 Mark of pyramid C**

## 4. CONCLUSIONS

Our project fulfilled goals with representing three-dimensional models of the major archaeological sites of the city of Tollan Cocotitlan. Augmented Reality technology has been used to present a model to show the marks of RA defined for the system and having the display city in its architectural boom, achieving user interactivity, in a nice and easy way to manipulate objects.

By means of our approach, it is possible to travel through archeology museums, exhibitions or in the same archaeological site as presented to the general public or as ancient cultures and civilizations had been developed.

To validate our application, we choose Xicocotitlan Tollan, that was one of the most important cities in the history of Mexico and served as the basis for the development of other cultures, as the Mayan culture.

AR places virtual objects in a real environment, allowing users to get a view of what is supplemented watching and with the possibility to transform these virtual objects, such as observing the virtual object from different perspectives views.

The aim of augmented reality is to set virtual objects of the real world, complementing what the user is watching and he can manipulate the virtual objects. In this case, Augmented Reality presents an interactive way to know the architecture of the Archeological Site Tollan, making a friendly system for the user to enrich the knowledge about this Culture.

## AKNOWLEDGEMENTS

The authors thank the school of Computing National Polytechnic Institute (ESCOM-IPN, México), for the economical support and the facilities provided for the development of this research work.

## Authors

**M. Sc. Marco Antonio Dorantes González**. Was born at Córdoba, Veracruz on 28 June, 1968. He had done his graduation in Electronics from ITO, Veracruz, México in 1990. After that he had completed his M. Sc. Degree in Computing in CINVESTAV in 1996 and M. Sc. of computing technologies in CIDETEC-IPN in 2008, research professor of ESCOM (IPN). He has been research Professor since 1996. He is interested in: Mobile Computing, Software Engineering, Data Bases. He has directed more than 70 engineering degree theses. Technical reviewer of interested areas books of publishers (McGraw Gill, Thompson, Pearson Education), He has participated in several research projects and has held some administrative positions in the IPN, also has experience in the industrial sector in the area of instrumentation and electronics; has done graduate studies in some fields, he has participated in several television programs and publications in scientific journals. 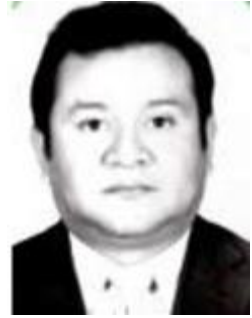

**M. Sc. Martha Rosa Cordero Lopez.** Was born at México D.F on 25 March, 1972. He had done his graduation in degree informatics from ITO, Veracruz, México in 1994. After that he had complete his Master Science Degree in Computing in CINVESTAV (IPN) in 1996, Master of computing technologies in CIDETEC-IPN in 2008, research professor of ESCOM (IPN) since 1995, her areas of interest are: Software engineering, Mobile Computing, Data Bases, affective computing, she has been director of in more than 70 theses to date, technical reviewer of interested areas books of publishes (McGraw Gill, Thompson, Pearson Education, among others). He has participated in various research projects and has held various administrative positions in the IPN also has experience in the private sector in the area of systems development; has done graduate studies in some areas, has been assistant manager of technology intelligence unit in the technological development of the IPN, has participated in some television programs and publications in scientific journals. 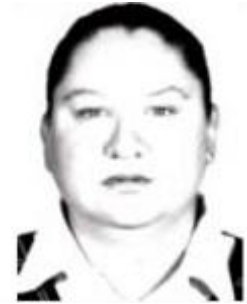